\documentclass{iau} 
\usepackage{graphicx}

\title[Cluster age and MPs] 
{The role of cluster age on the onset of multiple populations in stellar clusters }

\author[S. Martocchia]   
{S. Martocchia$^{1,2}$
}

\affiliation{$^1$European Southern Observatory, Karl-Schwarzschild-Stra\ss e 2, D-85748 Garching bei M\"unchen, Germany, \\ $^{2}$Astrophysics Research Institute, Liverpool John Moores University, 146 Brownlow Hill, Liverpool L3 5RF, UK\\ email: {\tt smartocc@eso.org} \\[\affilskip]
}

\pubyear{2019}
\volume{351}  
\jname{Star Clusters: From the Milky Way to the Early Universe}
\editors{A. Bragaglia, M.B. Davies, A. Sills \& E. Vesperini, eds.}
\begin{document}

\maketitle

\begin{abstract}
The origin of the chemical anomalies in star clusters is still an open question, although much effort has been employed both from a theoretical and observational point of view. The exploration of whether such multiple stellar populations are found based on certain properties of clusters 
has represented a compelling line of investigation so far. Here I report an overview of the results obtained from our latest surveys aimed 
at characterising the phenomenon of chemical variations in star clusters that are much younger with respect to the ancient globular clusters. The fundamental question we are asking is whether these abundance patterns are only restricted to the old massive clusters; and if not, is there a difference between young and old objects?
\keywords{star clusters, stars, multiple populations, chemistry, photometry, spectroscopy.}
\end{abstract}

\firstsection 
\section{Introduction}

All old and massive globular clusters (GCs) studied so far display
 star-to-star chemical anomalies in the form of anticorrelated 
 patterns among certain light elements. 
The most famous anti-correlations of such multiple stellar populations (MPs)
 are the C versus N (Cannon et al. 1998) and Na versus O (Carretta et al. 2009). 
MPs have been studied in many details (photometrically, spectroscopically, kinematically), 
however we are still far from understanding how they form (e.g. Bastian \& Lardo 2018). 
A powerful line of investigation that has been carried out so far is the exploration of the parameter space, 
i.e. whether MPs appear or not based on certain properties of the clusters. 
Here I would like to focus on two important properties of star clusters: their mass and age. 

Chemical anomalies have been searched and found in basically all massive and ancient GCs, in any enviroment, from the Milky Way to 
the Magellanic Clouds to the local dwarfs (see the recent compilation by Krause et al. 2016 and Bastian \& Lardo 2018 for a review). However, MPs have not been found 
in the so-called ``open'' clusters, which represent less dense, less massive objects ($\lesssim 10^4 M_{\odot}$, e.g. Bragaglia et al. 2012).
Hence, this led many to consider cluster mass as the key factor deciding the appearance of chemical anomalies in GCs. 
Many models trying to explain the origin of the chemical anomalies are indeed based on how massive a cluster is, although there are still many
observational evidence that is not explained. 
Since most of MPs studies were performed on old clusters ($>10$ Gyr), the natural question to ask is: are chemical anomalies only restricted to the ancient GCs? 

While our Galaxy lacks a population of massive ($>10^5 M_{\odot}$) stellar clusters with ages below $\sim$9$-$10 Gyr, our nearest galactic companions, the Large and Small Magellanic Clouds (LMC/SMC) host such clusters. How do we classify these clusters in terms of MPs studies? Do they share the same chemical patterns observed in the ancient globular clusters? Expanding the search for MPs towards different cluster ages is of extreme importance in order to potentially obtain new constraints for the formation mechanisms aiming at explaining the origin of chemical anomalies.

\section{Results}

\begin{figure}
\centering
\includegraphics[scale=0.62]{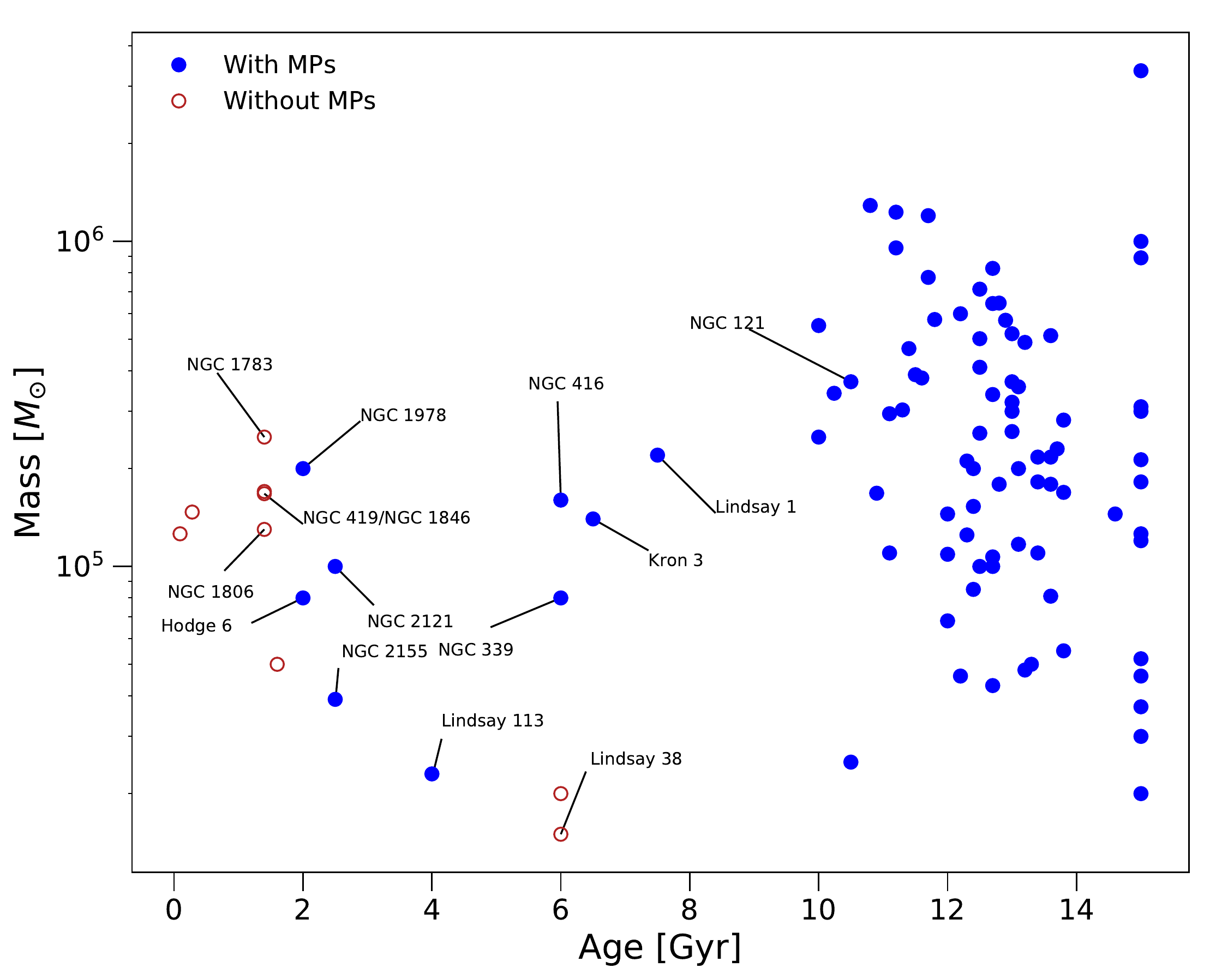}
\caption{Cluster mass versus cluster age diagram. The clusters from our sample are labeled. Blue filled (red open) circles represent sources with (without) multiple populations. See the compilation by Krause et al. 2016 and references therein. Adapted from Hollyhead et al. (2019).} 
\label{fig:agemass}
\end{figure}

To study MPs in young star clusters, we put together two surveys. Our Hubble Space Telescope (HST) photometric survey is composed of 13 star clusters both in the LMC and SMC. They are massive ($>$ a few times $10^4 M_{\odot}$) and they span a very wide range of ages (from $\sim$1.5 up to $\sim$10 Gyr). This is combined with a spectroscopic survey, which is composed by ESO-VLT FORS2 and XSHOOTER observations of four star clusters in the Magellanic Clouds (MCs), spanning ages from $\sim$2 up to $\sim$8 Gyr. Three new clusters will also be observed in September 2019 (age=1.5-2 Gyr) with FORS2. The goal of the survey is to search for a potential dependence on the onset of multiple population on cluster age, by looking at clusters that are as massive as the ancient GCs, but significantly younger. 
Our photometric technique consists in studying the red giant branch (RGB) stars in filters that are sensitive to N variations, as they encompass the NH molecular band (namely the HST F336W and F343N filters). Spectroscopically, we study molecular features in the UV such as CN and CH, sensitive to N and C spreads, respectively. With XSHOOTER we have spectra for two stars in the RGB of NGC 416, a $\sim$6 Gyr old cluster in the SMC and we are currently obtaining abundances for various elements such as N, C, Na and Mg. 

Results from our ongoing surveys have been published in Niederhofer et al. (2017a,b); Martocchia et al. (2017, 2018a,b, 2019, photometry) and in Hollyhead et al. (2017, 2018, 2019, spectroscopy). The first important result we obtained is the discovery of chemical anomalies as N variations in intermediate age clusters, i.e. aged $\sim$6-7.5 Gyr, both photometrically and spectroscopically. This corresponds to a redshift of formation for the MPs of $z=0.75$.

We then looked at the younger clusters. We did not find photometric evidence for MPs in clusters that are younger than 2 Gyr ($\sim1.5-1.7$ Gyr). 
However, we finally looked at MPs in star clusters in the age gap between 2 and 6 Gyr: two $\sim$2 Gyr old clusters, two aged $\sim$2.5 Gyr, and a 4 Gyr old one. We found MPs in the form of N variations in all of these clusters. We did not find evidence for chemical anomalies in Lindsay 38 though, a $\sim$ 6.5 Gyr old, but quite low mass cluster ($\lesssim 2\times 10^4 M_{\odot}$).

\section{Implications}
The main results from our surveys are shown in Figure 1, in the cluster age versus mass diagram. We keep finding MPs in intermediate age clusters, and for the first time, we find N variations in star clusters that are $\sim$2 Gyr old. This translates into a redshift of formation $z=0.17$. Thus, the first implication from our work is that chemical anomalies in the form of N spread are not restricted only to the ancient GCs. 
It does not seem to be the case for clusters that are younger than 2 Gyr, as they do not show MPs in the form of N spreads. The reason is still under investigation. By looking at RGB stars at different ages, we are also sampling stars with different stellar masses. RGB stars in a $\sim 2$ Gyr old population are less massive than 1.5$M_{\odot}$, while they become more massive than 1.5$M_{\odot}$ for populations younger than 2 Gyr. It would then be interesting to investigate whether this threshold is connected to other phenomena; indeed, stars below this mass threshold can be magnetically braked (Cardini \& Cassatella 2007). Also, the extended main sequence turnoff feature appears below this age threshold ($\sim$2 Gyr) in star clusters, but there are no hints of a correlation between this phenomenon and chemical anomalies to date. 

\begin{figure}
\centering
\includegraphics[scale=0.4]{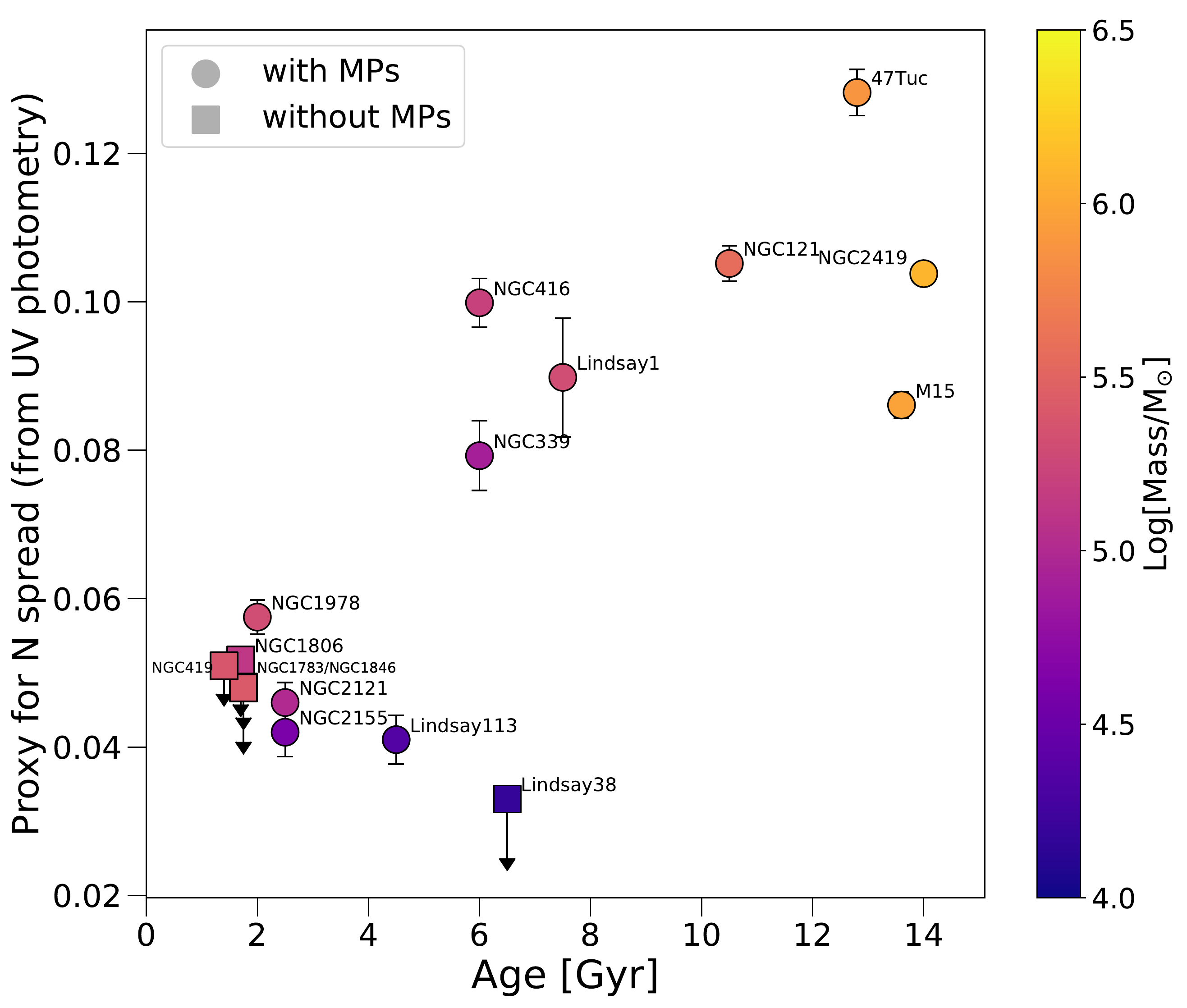}
\caption{Photometric proxy for N spread as a function of cluster age for the objects in our HST photometric survey plus M15, 47Tuc and NGC 2419. Data are colour coded by cluster mass. Figure adapted from Martocchia et al. (2019).} 
\label{fig:age}
\end{figure}

In our HST survey we also examined how the width of the RGB in the clusters varies as a function of cluster age. This is shown in Figure 2, where age is plotted against a quantity which is a proxy for N spread, from UV photometry. This Figure is adapted from Martocchia et al. (2019). A correlation between age and N spread is clearly visible, although there is a dependence on cluster mass as well. However, if we observe the 6-8 Gyr old clusters, they have a photometric spread around $\sim$0.09, while the younger clusters have a spread that is around $\sim$0.05. Hence, we observe that older clusters show larger abundance spreads compared to the younger clusters.

To conclude, in our surveys we started to characterise the young star clusters that we observe in the MCs, to gain new insights into the origin of the chemical variations in GCs. We observe chemical anomalies in the form of N spread in clusters older than $\sim$2 Gyr and we also observe that N variations increase as a function of cluster age. Future steps definitely involve more exploration of the parameter space of cluster properties, but most importantly, they involve the full chemical characterisation of young star clusters, as so far we have explored only N variations (and He to a certain extent, Chantereau et al. 2019, Lagioia et al. 2019).


\begin{thebibliography}{}

\bibitem[Bastian \& Lardo 2018]{bl18}
{Bastian N., Lardo C.} 2018,
\textit{ARA\&A}, 56, 83.

\bibitem[Bragaglia A. \etal\ (2012)]{bragaglia12}
{Bragaglia A., Gratton R. G., Carretta E., D’Orazi V., Sneden C., Lucatello S.} 2012,
\textit{ARA\&A}, 548, A122.

\bibitem[Cannon R. D. \etal\ (1998)]{cannon98}
{Cannon R. D., Croke B. F. W., Bell R. A., Hesser J. E., Stathakis R. A.} 1998,
\textit{MNRAS}, 298, 601.
 
 \bibitem[Cardini \& Cassatella (2007)]{cc07}
{ Cardini D., Cassatella A.} 2007,
\textit{ApJ}, 666, 393.

\bibitem[Carretta E.  \etal\ (2009)]{carretta09}
{Carretta E. et al.} 2009,
\textit{A\&A}, 505, 117.

\bibitem[Chantereau W.  \etal\ (2019)]{chant19}
{Chantereau W., Salaris M., Bastian N., Martocchia S.} 2019,
\textit{MNRAS}, 484, 5236.

\bibitem[Hollyhead K.  \etal\ (2017)]{h17}
{Hollyhead K. et al.} 2017,
\textit{MNRAS}, 465, L39.

 \bibitem[Hollyhead K.  \etal\ (2018)]{h18}
{Hollyhead K. et al.} 2018,
\textit{MNRAS}, 476, 114.
 
\bibitem[Hollyhead K.  \etal\ (2019)]{h19}
{Hollyhead K. et al.} 2019,
\textit{MNRAS}, 484, 4718.

\bibitem[Krause M. G. H.  \etal\ (2016)]{krause16}
{Krause M. G. H., Charbonnel C., Bastian N., Diehl R.} 2016,
\textit{A\&A}, 587, A53.

 
 \bibitem[Lagioia E.  \etal\ (2019)]{lagioia19}
{Lagioia, E. P., Milone, A. P., Marino, A. F., Dotter, A.} 2019,
\textit{ApJ}, 871, 140. 


\bibitem[Martocchia S.  \etal\ (2017)]{m17}
{Martocchia S., et al.} 2017,
\textit{MNRAS}, 468, 3150.
 
 \bibitem[Martocchia S.  \etal\ (2018)]{m18a}
{Martocchia S., et al.} 2018a,
\textit{MNRAS}, 473, 2688.

 \bibitem[Martocchia S.  \etal\ (2018)]{m18b}
{Martocchia S., et al.} 2018b,
\textit{MNRAS}, 477, 4696.

 \bibitem[Martocchia S.  \etal\ (2018)]{m19}
{Martocchia S., et al.} 2019,
\textit{MNRAS in press},  arXiv:1906.03273.

\bibitem[Niederhofer F.  \etal\ (2017)]{n17a}
{Niederhofer F., et al.} 2017a,
\textit{MNRAS}, 464, 94.
 
 \bibitem[Niederhofer F.  \etal\ (2017)]{n17b}
{Niederhofer F., et al.} 2017b,
\textit{MNRAS}, 465, 4159.


\end{thebibliography}
\end{document}